\PassOptionsToClass{table}{xcolor}
\documentclass[sigconf,screen,english,noacm]{acmart}

\usepackage[utf8]{inputenc}

\usepackage{amsmath,amsfonts}
\usepackage{algorithmic}
\usepackage{graphicx}
\usepackage{textcomp}
\usepackage{xfp}
\usepackage{xcolor}
\usepackage{colortbl}

\AtBeginDocument{}

\copyrightyear{2026}
\acmYear{2026}
\setcopyright{cc}
\setcctype{by}

\usepackage{url}

\definecolor{spaceorange}{HTML}{ED820E}

\usepackage{fontawesome}
\usepackage{tikz}
\usetikzlibrary{tikzmark,shapes.symbols}
\usepackage{makecell}
\usepackage{booktabs}
\usepackage{csquotes}
\usepackage[capitalize,nameinlink,noabbrev]{cleveref}
\let\T\texttt

\usepackage{microtype}

\usepackage[inline]{enumitem}
\newlist{inlist}{enumerate*}{1}
\setlist[inlist]{itemjoin={{, }},itemjoin*={{, and }},label=(\(\roman*\)),mode=boxed}

\usepackage{forest}
\useforestlibrary{edges}

\def\SWidth{3.25cm}%
\forestset{
   internal/.style={gray,font=\sffamily},
   marker/.style={rectangle split parts=1,minimum width=0cm},
   norm-ast-node/.style={
      Rect,
      outer sep=1pt,
      rounded corners=1pt,
      Soft,
      s sep=2pt,
      l sep+=2mm,
      rectangle split,
      rectangle split parts=2,
      font=\ttfamily,
   },
   l2r/.style={
      grow'=east,
      anchor = west,
      parent anchor = east,
      child anchor = west,
   },
   rot90/.style={
      rotate=90,grow=east,anchor=north,parent anchor=south,child anchor=north
   },
   norm-ast/.style={
      norm-ast-node,
      minimum width=\SWidth,
      l2r,
      line join=round,
   },
   o/.style={edge label={node[pos=.5,fill=white,inner sep=1.25pt,rounded corners=1pt,font=\footnotesize] {\strut #1}}},
   al-center/.style={
      before computing xy={s/.average={s}{siblings}}
   }
}
\def\NodeSurround{~~}%
\newsavebox\NodeTitleBox
\newcommand\Node[3][\SWidth]{\global\setbox\NodeTitleBox=\hbox{\NodeSurround#2\NodeSurround}\usebox\NodeTitleBox \nodepart{two} \edef\w{\ifdim\wd\NodeTitleBox<#1 #1\else\wd\NodeTitleBox\fi}\parbox{\w}{\smaller\sffamily#3}}%

\newlist{blocklist}{enumerate*}{1}
\setlist[blocklist]{itemjoin={{\space}},itemjoin*={{\space}},label=(\(\roman*\)),mode=boxed}
\newlist{orlist}{enumerate*}{1}
\setlist[orlist]{itemjoin={{, }},itemjoin*={{, or }},label=(\(\arabic*\)),mode=boxed}

\usepackage{wrapfig}

\usetikzlibrary{arrows.meta,fit,backgrounds,calc}
\tikzset{
   Soft/.style={line join=round,line cap=round},
   All Soft/.style={every path/.append style={Soft}},
   Blob/.style={
      draw=black,
      circle,
      outer sep=2pt,
      minimum size=1.8em
   },
   Blobs/.style={
      every node/.append style={Blob}
   },
   Use/.style={Blob},
   RectRounding/.style={rounded corners=3pt},
   Rect/.style={
      draw=black,
      rectangle,
      RectRounding,
      outer sep=2pt,
      minimum size=1.8em
   },
   Def/.style={Rect},
   Rects/.style={
      every node/.append style={Rect,minimum width=#1}
   },
   Rects/.default={1.5em},
   Link/.style={
      draw,
      Soft,
      rounded corners=2pt,
      -Kite
   },
   Links/.style={
      every path/.append style={Link}
   },
   comm/.style={rectangle,draw,text width=7.5mm,align=center,minimum height=5mm,font=\small\ttfamily,fill=lightgray!22!white},
   d/.style={comm,rounded corners=2pt,hyperlink node={v:def}}, 
   u/.style={comm,rounded corners=2.5mm,hyperlink node={v:use}}, 
   v/.style={comm,signal, signal to=east and west, gray, text=black, inner sep=-4pt, text width=5mm,fill=lightgray!22!white,hyperlink node={v:value}}, 
   F/.style={comm,draw=gray,rounded corners=2pt,fill=white,inner xsep=.5em,hyperlink node={v:fdef}},
   fc/.style={comm,double,rounded corners=2.5mm, outer sep=1.15pt,hyperlink node={v:call}},
   w-back/.style={fill=white,inner sep=1pt},
   T/.style={font=\footnotesize\ttfamily,text=gray},
   code/.style={font=\ttfamily},
   dfidn/.style={
      circle,darkgray,fill opacity=.925,xshift=-1mm,yshift=.15mm,fill=white,draw,minimum size=1.8em,scale=.8,inner sep=-1pt
   },
   klabel/.style={font=\scriptsize\sffamily, inner sep=1pt,text=black,above,execute at begin node={\strut}},
   olabel/.style={midway,klabel},
   pos at/.style={above #1=-.5mm,yshift=-.4\baselineskip},
   old/.style={opacity=.65},
   graph-frame-style/.style={very thick,rounded corners=1.5pt,draw,black},
   graph-cd/.style={gray,font=\scriptsize,line cap=round}, 
   graph-cd-in/.style={graph-cd,{Kite[scale=.7]}-},
   graph-cd-out/.style={graph-cd,-{Kite[scale=.7]}}
}

\tikzset{
    hyperlink node/.style={
        alias=sourcenode,
        append after command={
            let     \p1 = (sourcenode.north west),
                \p2=(sourcenode.south east),
                \n1={\x2-\x1},
                \n2={\y1-\y2} in
            node [inner sep=0pt, outer sep=0pt,anchor=north west,at=(\p1)] {\hyperlink{#1}{\XeTeXLinkBox{\phantom{\rule{\n1}{\n2}}}}}
        }
    }
}

\def\DefineEdgeType#1#2#3{%
   \expandafter\def\csname edge#1short\endcsname{#1}
   \expandafter\def\csname edge#1long\endcsname{#2}
   \expandafter\def\csname edge#1desc\endcsname{\def\s{\textit{source}\xspace}\def\t{\textit{target}\xspace}#3}
}

\usepackage{xspace}

\def\GetEdge#1#2{\csname edge#1#2\endcsname}

\DefineEdgeType{reads}{reads}{\s reads from \t}
\DefineEdgeType{def-by}{defined-by}{\s is defined by the \t}
\DefineEdgeType{calls}{calls}{\s calls the \t}
\DefineEdgeType{returns}{returns}{\s returns to \t}
\DefineEdgeType{def-on-call}{defines (on call)}{if the call occurs, \s~(argument) defines \t~(parameter)} 
\DefineEdgeType{arg}{argument}{\t vertex is an argument of the \s}
\DefineEdgeType{side-effect}{side effect (on-call)}{if the call occurs, the \s effects on \t}
\DefineEdgeType{nse}{non-standard evaluation}{\s causes a non-standard evaluation of \t}

\RequirePackage[
   print,
   fakeminted
]{lib/xlistings/xlistings}
\lstcolorlet{keywordA}{black}
\lstcolorlet{keywordB}{black}
\lstcolorlet{keywordC}{black}
\lstcolorlet{literals}{black!50!purple}
\lstcolorlet{numbers}{red!50!orange!50!cyan!80!black}
\usepackage{code-animation/code-animation}
\SetCodeAnimationFadeOutMixin{!8!lightgray}

\xlstDefineStyles{%
 {keywordA: \color{xlst@colors@lst@keywordA}\bfseries},%
 {keywordB: \color{xlst@colors@lst@keywordB}},%
 {keywordC: \color{xlst@colors@lst@keywordC}},%
 {keywordD: \itshape},%
 {numbers:  \color{xlst@colors@lst@numbers}\slshape},%
 {literals: \color{xlst@colors@lst@literals}\slshape},%
 {comments: \color{xlst@colors@lst@comments}},
 {basic:   \ttfamily},%
 {command:  \color{xlst@colors@lst@command}}%
}
\RequirePackage[tt=false, type1=true]{libertine}
\RequirePackage[varqu]{zi4}
\RequirePackage[libertine]{newtxmath}

\makeatletter
\appto\input@path{{lib/sopra-collection/sopra-listings/}}
\LoadLanguage{fse-R,ts}
\xlstsetmintedstyle{plain}

\AtBeginDocument{\xlstInlinePreBreak{}\xlstPreBreak{}}

\RequirePackage{tabularx}

\RequirePackage[detect-all]{siunitx}
\newcounter{NumberOfToolRows}
\providecommand\NumberOfToolRows{0}
\AtEndDocument{
   \immediate\write\@auxout{\string\gdef\string\NumberOfToolRows{\the\value{NumberOfToolRows}}}
}

\usepackage{multicol}

\usepackage{xfp}
\usepackage[breakall]{xurl}

\def\PrintResultOfThe#1#2#3#4{%
   \edef\result{\fpeval{round(#1,#2)}}%
   \edef\chk{\fpeval{\result=0?1:0}}%
   \ifnum\chk=1\relax
      \edef\numcmpres{\fpeval{1/(10^#2)}}%
      \qty{< \numcmpres}{#4}%
   \else
      #3\relax
   \fi
}

\newcommand*\ThePercent[3][2]{%
   \edef\fst{\fpeval{#2}}%
   \edef\snd{\fpeval{#3}}%
   \edef\result{\fpeval{\fst/\snd * 100}}%
   \PrintResultOfThe{\result}{#1}{\ifdim\result sp=100sp\relax\ifdim\fst sp=\snd sp\else\(\approx\)\fi\fi\qty{\fpeval{\result}}{\percent}}{\percent}%
}

\def\TestsDataflow{231}

\def\TestsSlice{158}

\def\DatflowTimeMs{576}
\edef\NumberOfTests{\noexpand\num[round-precision=0]{\fpeval{\TestsDataflow+\TestsSlice}}}

\tikzset{
    path image shift/.style={}, 
    path image/.style={path picture={\node at ([path image shift]path picture bounding box.center) {#1};}},
}
\newsavebox\CurrImgBox
\tikzset{ImageScales/.style={xscale=1.2,yscale=1},ImageStyle/.style={draw,thin,gray!10},ImageCornerRound/.style={rounded corners=6.75pt}}
\def\ImageWithRoundedCorners#1#2{%
   \savebox\CurrImgBox{\includegraphics[width=#1]{#2}}%
   \tikz{%
      \foreach \i/\o in {1/1,2/.5} {%
         \path[
            path image shift={(0pt,0pt)},
            path image={\includegraphics[width=#1]{#2}},
            ImageCornerRound,
            fading transform={ImageScales},
            ImageStyle,
            opacity=\o
         ] (0,0) rectangle (\wd\CurrImgBox,\ht\CurrImgBox);%
      }
   }%
}
\input{segments/_flowrview.tex}

\begin{document}
\def\OurTitle{Supporting the Comprehension of Data Analysis Scripts}
\title{\OurTitle}

\author{Florian Sihler}
\email{florian.sihler@uni-ulm.de}
\orcid{0000-0001-7195-7801}
\affiliation{%
  \institution{Ulm University}
  \country{Germany}
}

\author{Oliver Gerstl}
\email{oliver.gerstl@uni-ulm.de}
\orcid{0009-0007-5612-0780}

\affiliation{%
  \institution{Ulm University}
  \country{Germany}
}

\author{Lars Pfrenger}
\email{lars.pfrenger@uni-ulm.de}
\orcid{0009-0000-8166-7023}

\affiliation{%
  \institution{Ulm University}
  \country{Germany}
}

\author{Julian Schubert}
\email{julian.schubert@uni-ulm.de}
\orcid{0009-0007-1284-8008}
\affiliation{%
  \institution{Ulm University}
  \country{Germany}
}

\author{Matthias Tichy}
\email{matthias.tichy@uni-ulm.de}
\orcid{0000-0002-9067-3748}
\affiliation{%
  \institution{Ulm University}
  \country{Germany}
}


\def\blk#1{\ignorespaces}
\begin{abstract}
   A lot of research relies on data analysis scripts to process, clean, and visualize data. 
   However, recent studies show that these scripts are often hard to comprehend and maintain, hindering reproducibility and reuse, accompanied by a lack of tool support for handling such scripts.
   In this work, we focus on the R~programming language, addressing this problem by presenting \textit{flowR} as an extension for the common data analysis IDEs Positron and VS~Code.
   Alongside a previously presented static backward program slicer, flowR provides an overview of data analysis scripts, interactive graph visualizations, linting, and inline value annotations to support data analysts.
   \blk{Method} FlowR incrementally analyzes R~projects by intertwining interprocedural data- and control-flow analyses to build a comprehensive dataflow graph, incorporating R's dynamic and explorative features. 
   Additionally, flowR offers a plugin system and interfaces, allowing the integration of further analyses, such as new linting rules or custom visualizations.
   \blk{Results} Requiring an average of \qty{\DatflowTimeMs}{\milli\second} to calculate the full dataflow graph of real-world projects, this enables near real-time feedback.
   {\def\UrlFont{\itshape}The demonstration video is available at \url{https://youtu.be/hJzr-r-NmMg}. For the full source code and extensive documentation, refer to \url{https://github.com/flowr-analysis/flowr}. To try the docker image, use \T{\texttt{docker run -\null-rm -it eagleoutice/flowr}}.}
\end{abstract}



\maketitle
\def\UrlFont{\slshape}

\errorcontextlines=99999
\overfullrule=2cm

\section{Introduction}

Assume the role of a researcher who found an interesting study online that analyzes a dataset similar to their own.
According to recent studies, chances are high that the accompanying scripts are either non-reproducible, non-executable, or simply difficult to comprehend~\cite{trisovic_largescale_2022,DBLP:conf/msr/IslamAW24,sajuholtdirkmangroliyableier2025,DBLP:conf/msr/SihlerPSTDD24}.
In previous work~\cite{DBLP:conf/kbse/SihlerT24,sihler2025statically}, we presented \textit{flowR} as a dataflow analysis framework and backward program slicer for the R~programming language, which can help researchers understand and maintain R~scripts.
Building upon this, we now present a significant extension of flowR, by adding several features to its integration into common data analysis environments:
\begin{inlist}
   \item support for various notebook formats
   \item an overview of important analysis steps and script dependencies
   \item impact slicing to understand how inputs affect the remaining analysis
   \item reproducibility-oriented linting rules with quick-fixes
   \item hover-over values to understand variable and data shapes at specific program points
   \item responsive visualizations of, for example, the control-flow graph
   \item[]\!more
\end{inlist}.
These techniques support researchers and reviewers in comprehending, maintaining, and reusing R~scripts more effectively.
Additionally, these features also support researchers during the script development process, for example, by providing near real-time linting feedback and quick-fixes directly in the code editor.
FlowR's integration also provides several features to support tool developers in building new analyses on top of flowR like the query API.

Alongside its previously supported integrations into VS~Code, as a server, REPL, and Docker image, flowR now also works in the browser (e.g., on \href{https://vscode.dev}{vscode.dev}) and in \href{https://positron.posit.co/}{Positron}. Moreover, it ships with an easy-to-use interface for custom analyses and debugging and a sophisticated online documentation, which is continuously updated to stay in sync with the latest flowR~version.

After providing a brief overview of important related work in~\cref{sec:related-work}, we present all newly supported features of flowR in~\cref{sec:supported-features} alongside explanations on how to extend flowR with new analyses in \cref{sec:extending-flowr} and discuss the brief evaluation 
 in~\cref{sec:evaluation}. Finally, we conclude and outline future work in~\cref{sec:conclusion}.
\section{Related Work}\label{sec:related-work}
There are various tools and approaches to support developers and researchers in understanding and analyzing their programs. Language servers~\cite{gunasinghe2021language}, like the R~Language Server~\cite{rlanguageserver}, provide features such as code completion and go-to-definition functionalities. 
An\-alysis frameworks like SootUp~\cite{DBLP:conf/tacas/KarakayaSKBSLH24} for Java or CodeQL~\cite{codeql2026} offer static analysis to efficiently analyze programs for potential issues or vulnerabilities while linters like \textit{lintr}~\cite{lintr25} for~R, or \textit{ESLint}~\cite{eslint26} for JavaScript help identify and fix common issues based on rules.

More focused on data science, recent works explore analysis-specific validity checks~\cite{DBLP:conf/nsad/DolcettiCUZ24}, verifying statistical reporting~\cite{statcheck2020}, or expressing and verifying statistical assumptions within the analysis code~\cite{DBLP:journals/pacmse/TurcotteW25}.
However, so far related work is mostly unconnected to the problems mentioned in executability and reproducibility studies~\cite{DBLP:conf/msr/IslamAW24,trisovic_largescale_2022}. Moreover, these works do not focus on supporting comprehension, e.g., by incorporating the data science workflow as described by \citeauthor*{wickham2017r}~\cite{wickham2017r}.
With flowR, we fill this gap, as explained in more detail in~\cref{sec:supported-features}.
\section{New Features}\label{sec:supported-features}

This section presents the new features that \textit{flowR} now supports to help R~users comprehend, maintain, and develop R~scripts more effectively, on top of the previously presented techniques~\cite{DBLP:conf/kbse/SihlerT24}.
Every feature is presented alongside its integration into the extension, focusing on the benefits for end-users. For the underlying techniques and how tool developers can build upon them, please refer to \cref{sec:extending-flowr}.
By default, flowR uses a system-independent, \href{https://tree-sitter.github.io/tree-sitter/}{tree-sitter} based \href{https://r-lib.github.io/tree-sitter-r/index.html}{R~parser} and setup to analyze R~code, which allows it to work in the browser or without any R installation present. However, flowR can also make use of the R installation on the user's system.
\cref{fig:flowrview} provides a simplified overview of flowR.


\subsection{Notebook Support}\label{subsec:notebook-support}

Using flowR's plugin system, we provide built-in support for popular notebook formats: \href{https://github.com/flowr-analysis/flowr/blob/8ff867aaadab9ce92b4760d473cc734ae37bea9f/src/project/plugins/file-plugins/notebooks/flowr-analyzer-jupyter-file-plugin.ts#L14}{Jupyter Notebooks}, \href{https://github.com/flowr-analysis/flowr/blob/8ff867aaadab9ce92b4760d473cc734ae37bea9f/src/project/plugins/file-plugins/notebooks/flowr-analyzer-qmd-file-plugin.ts#L14}{Quarto}, and \href{https://github.com/flowr-analysis/flowr/blob/8ff867aaadab9ce92b4760d473cc734ae37bea9f/src/project/plugins/file-plugins/notebooks/flowr-analyzer-rmd-file-plugin.ts#L15}{R Markdown}.
This support includes parsing these formats, extracting the R~code from cells, and mapping analysis results to their respective locations in the notebooks.
This allows a seamless integration of flowR's features into notebook environments, which means that all subsequent and previously described features~\cite{DBLP:conf/kbse/SihlerT24} work in notebooks (or projects that include notebooks) as well.
Moreover, the information gained by these plugins are made available, allowing to easily build notebook-aware analyses on top of flowR (cf.~\cref{sec:extending-flowr}). Users can also add support for custom notebook formats by implementing their own plugin.

\subsection{Impact Slicing}\label{subsec:impact-slicing}

{\columnsep=.785em\intextsep=0pt
Corresponding to the previously presented backward slicing~\cite{DBLP:conf/kbse/SihlerT24}, which allows users to understand which parts of the code influence a specific program point, we now also support forward or \emph{impact slicing}.
Given an element in the program, an impact slice contains all code that is influenced by this element (also called a \enquote{forward slice}).
This feature helps users to understand how specific or multiple inputs affect the remaining analysis, for example, by showing all expressions that operate on a specific dataset.
Similar to the drastic reductions we found with backward slices~(to around \qty{13.5}{\percent}, cf.~\cite{sihler2025statically}) the average forward slices on real-world sources reduces the project to just around~\qty{13}{\percent} of the original code, making it easier to comprehend the effects of specific inputs. Please note that in contrast to backward slicing, the forward slices presented by flowR are usually not executable, as they, by-design, omit other inputs that may be required for the data analysis.
Moreover, combining backward and forward slicing (a process known as \enquote{chopping}~\cite{DBLP:conf/sigsoft/JacksonR94}) allows to reduce the analysis to only the parts that connect specific inputs and outputs. This helps to answer questions like \enquote{How is the data from this CSV file processed to produce this plot?} more easily.
\par}


\subsection{Dependency Overview}\label{subsec:dependency-overview}

{\columnsep=.785em\intextsep=0pt\hyphenpenalty=1000\relax
\begin{wrapfigure}{r}{.605\linewidth}
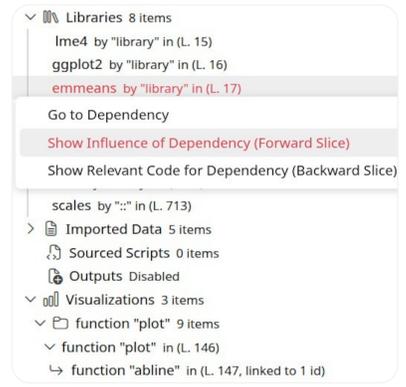

   \centering
   \ImageWithRoundedCorners{\linewidth}{img/dep-view}\\[-4pt]
   \textcolor{darkgray}{\tiny Generated for the analysis from \hypersetup{allcolors=.}\citeauthor{scriptexample}~\cite{scriptexample}}\vspace{\dimexpr-1pt-\baselineskip\relax}
   \caption{Dependency View}\label{fig:dep-view}
\end{wrapfigure}
To help researchers understand the structure of a given data analysis, we provide an automatically generated and updated overview of all script dependencies and outputs~(cf.~\cref{fig:dep-view} to the right).
This overview categorizes important steps of the analysis, such as data loading and visualizations, alongside all used libraries and the produced outputs.\\
\hspace*{\parindent}Each item is automatically linked to its location in the project so a user can quickly navigate to it by clicking on the respective element.
Additionally, users can directly inspect the corresponding backward slice (i.e., all code that influences the respective element~\cite{DBLP:conf/kbse/SihlerT24}) or the impact of an element~(\cref{subsec:impact-slicing}) as shown by the context menu open in \cref{fig:dep-view}.
This makes it easy to answer questions such as \enquote{What are the visualizations produced by this script?} or \enquote{Which parts of the analysis depend on this input file?}.
If an element is semantically connected to another one---for example, a plot function like \textit{abline} in \cref{fig:dep-view} that draws to the same figure created by the \textit{plot} functions---the view will group them automatically.
This allows to see related parts together, even if they are separated in the code.
The overview is highly configurable, e.g., by allowing to disable elements such as \textit{Outputs} in \cref{fig:dep-view} or adding custom categories.
\par}


\subsection{Linting Integration}\label{subsec:linting-integration}
\begin{figure}[H]
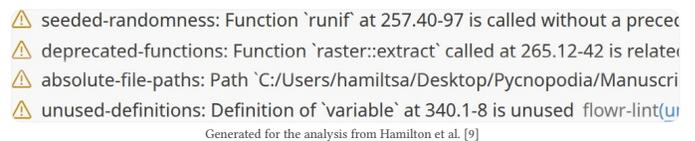

\tikzset{ImageCornerRound/.style={rounded corners=3.75pt}}
~\clap{\ImageWithRoundedCorners{\dimexpr\linewidth+11.5pt}{img/linter-results}}~\\[-5pt]
\textcolor{darkgray}{\tiny Generated for the analysis from \hypersetup{allcolors=.}\citeauthor{linterexample}~\cite{linterexample}}\vspace{-\baselineskip}
\caption{Excerpt of linting results}\label{fig:linting-results}\vspace*{-.55\baselineskip}
\end{figure}

flowR includes a set of ten different and highly configurable \href{https://github.com/flowr-analysis/flowr/wiki/Linter}{linting rules} designed to identify common issues that hinder the executability and reproducibility of R~scripts as exemplified in \cref{fig:linting-results}.\vspace*{-.55\baselineskip}

\noindent These rules check for various issues such as 
\begin{orlist}
   \item absolute paths which are usually not portable to different systems
   \item invalid file paths indicating missing or misplaced inputs
   \item accessing non-existent columns in data frames hinting at wrong or incomplete data processing
   \item using a random number source without initializing it with a fixed seed
\end{orlist}, as indicators for non-executable or non-reproducible code.
Many of these rules also provide quick-fixes that automatically resolve the identified issues, e.g., by replacing absolute paths with relative ones based on the project's root directory or by setting a fixed seed for random number generators.
These quick-fixes also support and respect R's dynamic features like relative directory changes and path constructions~\cite{sihler2025statically}.
The linting results are presented in a dedicated view within the extension, allowing users to quickly navigate to and fix issues in their code or try to automatically resolve them using the provided quick-fixes.
The presented linting results can be silenced individually and linting rules can be configured extensively to fit the user's needs.
Additionally, using \href{https://github.com/flowr-analysis/flowr/wiki/Linter}{the API}, tool developers can easily extend the linter with new rules or update existing ones. 

\subsection{Hover-Over Value Provider}\label{subsec:hover-over-value-provider}

{\columnsep=.785em\intextsep=0pt
\begin{wrapfigure}[8]{r}{.525\linewidth}
\vspace*{-.35\baselineskip}\lstfs{7}
\begin{minted}[escapeinside=||]{R}
# ...
coln <- "id"
x <- a                   %>%
   mutate(level=score^2) %>%
   left_join(b, b:c:y=|\tikzmarknode{coln}{\texttt{coln}}|) %>%
   |\tikzmarknode{select}{\texttt{select}}|(-age)
\end{minted}
\begin{tikzpicture}[overlay,remember picture,tt/.style={
   font=\scriptsize,
   fill=gray!2,
   rounded corners=3pt,
   draw=lightgray,thin,
   inner sep=3pt,
   text=darkgray,
   drop shadow
}]
\draw[Circle-Kite,lightgray,font=\footnotesize] (coln.north) to[out=70,in=0] ++(-6pt,4.75mm) node[left,tt] {Value: \texttt{"id"}};
\draw[Circle-Kite,lightgray,font=\footnotesize] ([xshift=-2mm]select.south) to[out=-110,in=180] ++(6pt,-5mm) node[right,tt,text width=3.5cm] {Returns a data frame with 4 rows, and known columns: \texttt{foo}, \texttt{score}, \texttt{level}, \texttt{id}.};
\end{tikzpicture}
\caption{Hover-over values}\label{fig:hover-over-value-provider}
\end{wrapfigure}
To help users understand the values of variables and data shapes at specific program points, flowR provides value information in the code editor. 
When hovering over a variable or expression in the code editor, a tooltip appears showing the computed value or data shape at that point in the program as exemplified in~\cref{fig:hover-over-value-provider}.
These values are also used internally by other analyses, e.g., to determine whether a column access could fail due to a missing column in a data frame or whether a given piece of code is dead because its condition is set to be constantly \texttt{FALSE}~\cite{sihler2025statically}.
While the idea follows the approach of various teaching tools and debuggers, which usually indicate the concrete values of variables~\cite{DBLP:conf/icse/GuptaRC22,DBLP:conf/uist/KangG17}, we also provide abstracted descriptions that, for example, summarize the shape of data frames even if the concrete values are not known statically.
Providing these values also helps while writing new analyses in~R, as they provide quick feedback on the expected values of variables without needing to run the code (or even converting it to an executable form).
\par
}
All these values are computed using flowR's interprocedural static analysis and fixpoint solver and are available throughout the analysis process~(cf.~\cref{sec:extending-flowr}).

\subsection{Responsive Graph Views}\label{subsec:responsive-graph-views}
FlowR provides a various graphs and views to help users understand the code but also to support tool developers in building new analyses: \begin{inlist}
   \item a normalized AST
   \item dataflow graph~(DFG)
   \item call graph~(CG)
   \item control flow graph~(CFG)
\end{inlist}. All of these can be directly visualized using \href{https://mermaid.live/}{mermaid.live} and shown within the editor. \cref{fig:compact-cfg} below exemplifies the presentation of a simplified CFG:\vspace*{-.65\baselineskip}

\begin{figure}[H]
\centering
\null\hfill\begin{minipage}{.35\linewidth}
\begin{minted}{R}
x <- 0
while(x < 20) {
   x <- x + 1
}
\end{minted}
\end{minipage}\hfill\begin{minipage}{.4\linewidth}
\scalebox{.75}{\begin{tikzpicture}[block/.style={Rect,font=\ttfamily,minimum width=1.75cm}]
\node[block] (bb1) {x <- 0};
\node[block,rounded corners=8pt,below=3mm] (bb2) at (bb1.south) {x < 20};
\node[block,right=3mm] (bb3) at (bb2.east) {x <- x + 1};
\node[fill,circle,double,inner sep=3pt,below=3mm] (bb4) at (bb2.south) { };
\draw (bb4.center) circle[radius=6pt];
\node[fill,circle,double,inner sep=3pt,right=3mm] (bb0) at (bb1.east) { };
\draw[-Kite,gray,densely dotted] (bb1.south) -- (bb2.north);
\draw[-Kite] (bb2.east) to[edge node={node[above] {\(\top\)}}] (bb3.west);
\draw[-Kite,gray,densely dotted] (bb3.north) to[bend right=15] ([xshift=4mm]bb2.north);
\draw[-Kite] (bb2.south) to[edge node={node[right] {\(\bot\)}}] (bb4.north);
\draw[-Kite,gray,densely dotted] (bb0.west) -- (bb1.east);
\end{tikzpicture}}
\end{minipage}\hfill\null\vspace*{-.5\baselineskip}
\caption{Example Code and the Compact CFG}\label{fig:compact-cfg}
\end{figure}
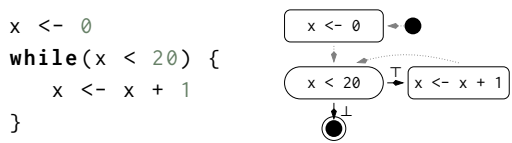\vspace*{-.65\baselineskip}

\noindent These views automatically respond to content changes and selections in the code editor. For example, by only showing the relevant parts of the graph for the currently selected code.
Additionally, these views provide various configurations to adapt the presentation of the graphs to the user's needs, e.g., by allowing to toggle between a compact and detailed view of the CFG or automatically removing dead code from the graphs. In general, these views can be opened on demand with the command palette of the editor, e.g., using the command \textit{\enquote{flowR: Show Dataflow Graph}}.

While the inspection of a \href{https://github.com/flowr-analysis/flowr/wiki/Dataflow-Graph}{dataflow graph} or the \href{https://github.com/flowr-analysis/flowr/wiki/Normalized-AST}{normalized AST} may be only interesting for tool developers~\cite{DBLP:conf/wcre/WeckT16}, the \href{https://github.com/flowr-analysis/flowr/wiki/Dataflow-Graph\#perspectives-cg}{call graph} and especially the simplified \href{https://github.com/flowr-analysis/flowr/wiki/Control-Flow-Graph}{control-flow graph} are helpful to provide an overview of a program's structure and possible execution paths.
Moreover, they also serve as the underlying structures for all other analyses~(cf.~\cref{sec:extending-flowr})

\begin{figure*}
\begin{minipage}[c]{\dimexpr0.775\linewidth-1.25em}
\centering\resizebox*{\dimexpr\linewidth}!{\usebox\FlowRview}
\end{minipage}\hfill\begin{minipage}[c]{.22\linewidth}
   \small\slshape\color{darkgray}The {\href{https://github.com/flowr-analysis/flowr/wiki/Analyzer\#Builder_Configuration}{Flowr\-Analyzer\-Builder}} class serves as the main entry point to configure and construct analyses with as well as on-top of \textit{flowR}.\medskip
   
   The resulting {\href{https://github.com/flowr-analysis/flowr/wiki/Analyzer\#overview-of-the-analyzer}{Flowr\-Ana\-lyz\-er}} orchestrates the complete analysis process, providing methods such as \texttt{dataflow} or \texttt{query} to calculate the results on-demand and provide full access to all components indicated to the left.
   During the analysis, \href{https://github.com/flowr-analysis/flowr/wiki/Analyzer\#Plugins}{plugins} provide support for various file-types, modify the project discovery process, and enrich the analysis with additional information. The \href{https://github.com/flowr-analysis/flowr/wiki/Analyzer\#Context_Information}{context} can be inspected to retrieve additional information about the current analysis state (see \cref{sec:extending-flowr}).
\end{minipage}
\caption{A simplified view of flowR's processes and results. All important steps are hyperlinked to their documentation.}\label{fig:flowrview}
\end{figure*}

\section{Extending flowR}\label{sec:extending-flowr}

Besides the extension with its new features as presented in~\cref{sec:supported-features},
flowR offers a TCP and WebSocket server interface~\cite{DBLP:conf/kbse/SihlerT24} as well as a \href{https://github.com/flowr-analysis/flowr/wiki/Interface\#using-the-repl}{read-eval-print loop~(REPL)} directly accessible from the \href{https://hub.docker.com/r/eagleoutice/flowr}{Docker image}: \T{\texttt{docker run -\null-rm -it eagleoutice/flowr}}. Moreover, flowR is available as an \href{https://github.com/flowr-analysis/flowr-r-adapter}{\textit{R}} and an \href{https://www.npmjs.com/package/@eagleoutice/flowr}{\textit{npm package}}, the latter of which we use in the extension presented in this work. FlowR is mostly implemented in TypeScript with parts in~R.

In this section, we present flowR's core pipeline architecture~(\cref{subsec:pipeline-architecture}) and the project analyzer~(\cref{subsec:project-analyzer}) from a tool developer's perspective. For the underlying techniques, please refer to our previous work~\cite{sihler2025statically} and the \href{https://github.com/flowr-analysis/flowr/wiki/Analyzer}{extensive documentation}.

\subsection{Pipeline Architecture}\label{subsec:pipeline-architecture}

For each R~source to be analyzed, flowR follows a multi-step \href{https://github.com/flowr-analysis/flowr/wiki/Core#pipelines-and-their-execution}{analysis pipeline}
which can be freely configured and extended, but in general consists of three main steps:
\begin{enumerate}[leftmargin=*]
   \item \href{https://github.com/flowr-analysis/flowr/wiki/Core\#parsing}{\textit{Parsing}}, to obtain the abstract syntax tree~(AST).
   \item \href{https://github.com/flowr-analysis/flowr/wiki/Core\#normalization}{\textit{Normalization}}, to transform the AST into a normalized and version-independent representation of the R~code.
   \item \href{https://github.com/flowr-analysis/flowr/wiki/Core\#dataflow-graph-generation}{\textit{Data- and Control-flow analysis}}, to compute the dataflow graph alongside the control-flow based on the normalized AST.
\end{enumerate}
To parse and normalize the source, flowR provides two back-ends: \begin{inlist}
   \item a tree-sitter based parser using a specialized \href{https://r-lib.github.io/tree-sitter-r/index.html}{R~grammar} without any external dependencies
   \item an R~based parser using R's built-in \href{https://www.rdocumentation.org/packages/base/versions/3.6.2/topics/parse}{\texttt{parse}} function but requiring R to be installed on the system
\end{inlist}. Both back-ends produce the same \href{https://github.com/flowr-analysis/flowr/wiki/Normalized-AST}{normalized AST} representation which then serves as the input for the subsequent interprocedural data- and control-flow analysis.
Using the unique identifiers of every node in the normalized AST, all subsequent analyses can then relate their results back to the original source code.
Based on the dataflow graph produced by the last step, various static analyses can be performed, e.g., \href{https://github.com/flowr-analysis/flowr/blob/d9dd97d72818c9875085eb643f9aeaf034312b57/src/slicing/static/static-slicer.ts}{slicing}, \href{https://github.com/flowr-analysis/flowr/blob/d9dd97d72818c9875085eb643f9aeaf034312b57/src/dataflow/graph/call-graph.ts}{call graph generation}, or \href{https://github.com/flowr-analysis/flowr/wiki/Dataflow-Graph\#resolving-values}{value resolution}. These analyses then directly contribute to the features presented in \cref{sec:supported-features}.

To add new or replace existing steps, users can simply implement the \href{https://github.com/flowr-analysis/flowr/blob/d9dd97d72818c9875085eb643f9aeaf034312b57/src/core/steps/pipeline-step.ts\#L70}{\textit{IPipelineStep}} interface, combine them with other steps, and pass the created pipeline to the \href{https://github.com/flowr-analysis/flowr/blob/main/src/core/pipeline-executor.ts\#L97}{\textit{PipelineExecutor}}. 

\subsection{Project Analyzer}\label{subsec:project-analyzer}
The project analyzer is the main API of flowR's analysis pipeline and it orchestrates the complete analysis process. The underlying architecture is depicted in \cref{fig:flowrview}, the pipeline as explained in \cref{subsec:pipeline-architecture} is highlighted in red.
Given an R~project, flowR first discovers, loads, and prepares all relevant files using its \href{https://github.com/flowr-analysis/flowr/wiki/Analyzer\#Plugins}{plugin system} including build specifications if present and then analyzes the contained R~sources using the configured pipeline. 
\paragraph{Plugins} Plugins are automatically applied during the analysis process to provide support for \begin{inlist}
   \item discovering files in a project
   \item loading and parsing specific file types
   \item enriching the context of a project
   \item identifying the optimal loading order of files
\end{inlist}.
New plugins can be created by building on the \href{https://github.com/flowr-analysis/flowr/blob/d9dd97d72818c9875085eb643f9aeaf034312b57/src/project/plugins/flowr-analyzer-plugin.ts#L87}{\textit{FlowrAnalyzerPlugin}} base class or one of its specializations.

\paragraph{Analysis Example} With the project analyzer~API, obtaining the dataflow graph of a project looks like this:
\begingroup\lstfs{8}
\begin{minted}[
   morekeywords={const},
   morekeywords={[2]{FlowrAnalyzerBuilder,setEngine,build,addRequest,dataflow}},
   keywordstyle={[2]\color{darkgray}\itshape}
]{typescript}
const analyzer = await new FlowrAnalyzerBuilder()
      .setEngine('tree-sitter')
      .build();
analyzer.addRequest('file:///path/to/project');
const df = await analyzer.dataflow();
\end{minted}
\endgroup
The builder automatically registers a set of \textit{\href{https://github.com/flowr-analysis/flowr/blob/d9dd97d72818c9875085eb643f9aeaf034312b57/src/project/plugins/flowr-analyzer-plugin-defaults.ts}{default plugins}} which provide support for notebooks as presented in~\cref{subsec:notebook-support} and other file types commonly found in R~projects.
To inspect a project, flowR provides its \href{https://github.com/flowr-analysis/flowr/wiki/Query-API}{query~API}. With it, we can obtain the value of a variable using queries like the \textit{\href{https://github.com/flowr-analysis/flowr/wiki/Query-API\#resolve-value-query}{resolve value query}}. This query is used by the the hover-over value provider as presented in \cref{subsec:hover-over-value-provider}. As a quick example, the following code uses the \textit{\href{https://github.com/flowr-analysis/flowr/wiki/Query-API\#dependencies-query}{dependencies query}} which is also used to provide the dependency overview presented in \cref{subsec:dependency-overview} to list all libraries used in the analyzed project:
\begingroup\lstfs{8}
\begin{minted}[
   morekeywords={const},
   morekeywords={[2]{log,query,dependencies,value,derivedVersion}},
   keywordstyle={[2]\color{darkgray}\itshape},
   escapeinside=||,
]{typescript}
const q = await analyzer.query([
   { type: 'dependencies' }
]);|\smallskip|
for(const lib of q.dependencies.library) {
   console.log(lib.value, lib.derivedVersion);
}
\end{minted}
\endgroup
The project API is type-safe, including the results of the queries, which allows for an easier exploration using autocompletion.

\section{Evaluation}\label{sec:evaluation}

We evaluate the performance of flowR's dataflow analysis following the ACM SIGPLAN~\cite{acmSigplanGuidelines} and SIGSOFT Empirical Standards~\cite{acmSigsoftGuidelines} using a dataset of \num{4230} real-world R~scripts provided as supplements to publications~(cf.~\cite[Section~7]{sihler2025statically}).
Analyzing each project, we measure the time required to build the inter-procedural dataflow graph as well as the size of the resulting graph. For the analysis, any multithreading and caching is disabled to reflect a bottom line for flowR's cold-start performance.

On average, we require \qty{115}{\milli\second} to parse and normalize a project using the tree-sitter back-end~(cf.~\cref{subsec:pipeline-architecture}). Based on the normalized AST, we build the dataflow graph in \(\approx{}\)\qty{525}{\milli\second} on average, resulting in a total analysis time of \(\approx{}\)\qty{640}{\milli\second} per project. Please note that these numbers are biased by outliers, with a median total time of just \qty{\fpeval{51+41+159}}{\milli\second}.\footnote{Using a Linux~PC with a~\qty{5}{\giga\hertz}~Intel i9-9900K CPU and sufficient memory.}
The average dataflow graph contains~\num{1733} vertices and~\num{3738} edges, with a median size of~\qty{213}{\kilo\byte}~\cite{sihler2025statically}.
All other capabilities presented in \cref{sec:supported-features} only require \qtyrange{10}{100}{\milli\second} with minor outliers in specific linting rules like the access validation of data frames. By parallelizing these analyses in practice, executing queries lazily and caching, flowR can provide near real-time feedback, with an average of around a second for the complete analysis of a project including the subsequent analyses.

\section{Conclusion}\label{sec:conclusion}

In this work, we presented a significant extension to flowR, a sophisticated program analysis framework for the R programming language.
As an important next step, we plan to evaluate the usability and usefulness of flowR's new features in a user study with researchers from data science domains.

\begin{acks}
   This work was supported by the \grantsponsor{dfg}{German Research Foundation (DFG)}{https://gepris.dfg.de/gepris/OCTOPUS}: \textit{\grantnum{dfg}{504226141}}.
\end{acks}

\bibliographystyle{ACM-Reference-Format}
\bibliography{references}

\appendix
\section{Demonstration Walkthrough}

This appendix contains the walkthrough of our demonstration video. For the required information on the tool, please refer to \cref{sec:tool-information}.
The walkthrough is organized chronologically alongside the different parts in the video.

\subsection{The VS Code Extension}

The extension is readily available on the \href{https://marketplace.visualstudio.com/items?itemName=code-inspect.vscode-flowr}{VS~Code Marketplace} and can be installed like any other extension. For example, by searching for \enquote{R Code Analyzer (flowR)} in the extensions tab of VS~Code and clicking \textit{Install}. This should also work in the browser on \href{https://vscode.dev}{vscode.dev}, even though the web-version of VS~Code has some limitations on certain browsers so we recommend using \href{https://www.google.de/intl/de/chrome/}{Google Chrome}.
With the default configuration, flowR and all of the following steps work out of the box.
Additionally, explanations of the respective features alongside GIFs that explain how to access them in the extension are part of the \href{https://github.com/flowr-analysis/vscode-flowr}{extension's README}. 
To have a concrete example to test whether the extension is working correctly, we use the following synthetic example script:

\xlstsetmintedstyle{plain number}
\begin{minted}[deletekeywords={t,c},escapeinside={|@}{@|},morekeywords={filter,t.test,ggplot,geom_count,aes}]{R}
|@\label{ggplot2}@|library(ggp:c:lot2)

|@\label{read.csv}@|da:c:ta <- read.csv("/data/data.csv")
min_age <- 42
b:c:y_age <- da:c:ta |>
|@\label{dplyr}\label{filter}@|    dplyr::filter(age >= min_age)

|@\label{ggplot}@|ggplot(b:c:y_age, aes(x=age, y=m)) + 
|@\label{geom-count}@|    geom_count()
\end{minted}

\paragraph{The Dependency View~(\cref{subsec:dependency-overview}, \href{https://raw.githubusercontent.com/flowr-analysis/vscode-flowr/refs/heads/main/media/gif/show-dependencies-opt.gif}{GIF})}

After installing the flowR extension, the sidebar features a new icon of a flower: clicking it opens the dependency view, which, whenever an R~project is opened, shows an overview of the dependencies as presented in \cref{subsec:dependency-overview}. Alternatively, the command \enquote{Focus on Dependencies View} can be used to open the view.
The view updates automatically, by default based on the size of the project either on every change or in a fixed interval.
Clicking on the entries navigates to the respective position in the code where the dependency is introduced.
To inspect the backward slice~\cite{DBLP:conf/kbse/SihlerT24} and the impact slice~(\cref{subsec:impact-slicing}) of the respective dependency, the context menu can be accessed by right clicking on an entry (also shown in \cref{fig:dep-view}).

Using the example code introduced above, the dependency view shows two loaded libraries: \begin{enumerate}
   \item \texttt{ggplot2} in line~\ref{ggplot2}, loaded with the \texttt{library} function
   \item \texttt{dplyr} in line~\ref{dplyr}, loaded with the \texttt{::} operator
\end{enumerate}
Additionally, it shows one imported dataset with the \enquote{read.csv} function in line~\ref{read.csv} and a single visualization created with \texttt{ggplot} in line~\ref{ggplot} with one linked function call to \texttt{geom\_count} (as it draws to the same plot).

\paragraph{Impact Slicing (\cref{subsec:impact-slicing})}
By right clicking on an entry in the dependency view, the context menu allows to inspect the impact slice of the respective dependency. For example, the impact slice of the dataset imported in line~\ref{read.csv} includes all lines of the script but the library import and the assignment to \texttt{min\_age}, as all other expressions depend on the data.
Alternatively, an impact slice can be produced for any expression in the code by right clicking on the element and using the \enquote{Show Influence\ldots} and \enquote{Toggle Influence\ldots} commands in the context menu.
Any active highlighting can be cleared with the \enquote{flowR: Clear Current Slice Presentation} command or the shortcut~\texttt{Alt+C}.

Please note that currently, flowR does not (sensibly) support forward slicing on library imports, because flowR does not ship with the known export signatures of all R~packages yet~\cite{sihler2025statically}.

\paragraph{Accessing the Linter (\cref{subsec:linting-integration})}
While the linter is enabled by default, running in the background with similar update strategies as the dependency view, it can also be triggered manually using the command \enquote{flowR: Code Quality Analysis (Linter)}.
Spotted problems are highlighted directly in the code editor (if VS~Code is not configured otherwise) and also appear in the \textit{Problems}~tab.

For the example code, the linter spots only one issue: the hard-coded path in the \texttt{read.csv} call. However, playing around with the code, e.g., by adding an unused variable or calling a deprecated function such as \textit{filter\_all} should automatically trigger new linter warnings.
For an unused assignment like \bIndexR{x <- 2}, the linter also provides a quick-fix which removes the expression.\footnote{By default, the linter probably does not suggest a quick-fix for the absolute path as if you just paste the code into VS~Code without opening a regular project, flowR has no notion of the project root directory.}

\paragraph{Hover-Over Value Provider (\cref{subsec:hover-over-value-provider}, \href{https://raw.githubusercontent.com/flowr-analysis/vscode-flowr/refs/heads/main/media/gif/value-info.gif}{GIF})} 
By hovering over either occurrence of \texttt{min\_age} in the example code, flowR shows a hover-over tooltip with the inferred value~\texttt{[42L, 42L]} as an interval to indicate that the variable holds the integer value~42.
Without knowing the contents of the \textsc{csv} file, flowR cannot infer sensible shapes for the data frames~\texttt{data} and \texttt{by\_age} (also due to R's semantics based on whether \textit{age} is a column in the data frame or not).
However, if you supply a \textsc{csv} file with (at least) the columns \texttt{age} and \texttt{m} at the requested path, flowR can infer more precise shapes for the data frames and, consequently, also for the visualization.

\paragraph{Accessing Graph-Views (\cref{subsec:responsive-graph-views}, \href{https://raw.githubusercontent.com/flowr-analysis/vscode-flowr/refs/heads/main/media/gif/show-df-graph-opt.gif}{GIF})}
All graph-views can be accessed via the command palette. For example, by searching for \enquote{flowR: Show Dataflow Graph} and selecting the command.
This opens a new tab in VS~Code showing the dataflow graph alongside a toolbar at the bottom of the view that allows to configure and fine-tune the graph visualization.\footnote{Please note, that for large projects detailed views such as the dataflow graph might take some time to load and render due to our reliance on \href{https://mermaid.live/}{mermaid.live}. If the extension rejects the graph for being too large, you can try to increase the default value of \texttt{vscode-flowr.style.mermaidMaxTextLength} in the extension settings.}
While the configuration options differ between the graph views (e.g., control-flow graphs can be shown with various simplifications selected or not), they all share common options such as automatically reducing the graph to only show or highlight currently selected elements in the code editor.
Additionally, every graph view shows links to the corresponding documentation and to the underlying \href{https://mermaid.live/}{mermaid} graph in the upper right corner.

\paragraph{Notebook Support (\cref{subsec:notebook-support})}
All of the aforementioned features also work in R~notebooks, as explained in \cref{subsec:notebook-support}. However, you have to make sure that VS~Code also labels the file as an R~notebook or R~code in general, as flowR does not activate if, for example, VS~Code thinks the file is a plain text file. Moreover, please note, that currently, flowR offers no support for the non-trivial execution model of R~notebooks, i.e., the analysis assumes that all code cells are executed in order from top to bottom without skipping or re-executing cells~\cite{sihler2025statically}.

\paragraph{Configuration Options}
Searching for \textit{vscode-flowr} in the settings (or using the settings icon shown in the title bar of the dependency view) opens the configuration options of the extension.
These allow, among others, to configure the update strategies of the analyses, enable/disable specific analyses, and to configure advanced options such as the port used to communicate with the flowR server.

\subsection{Using flowR in Positron and the Browser}
Using \href{https://open-vsx.org/extension/code-inspect/vscode-flowr}{flowR's Open VSX package} in \href{https://positron.posit.co/}{Positron} changes nothing compared to the VS~Code extension, providing all features in a similar manner. Yet, the default configuration of the IDEs might differ, so it may be that linter results are not shown automatically because the corresponding tab is hidden by default.

In \href{https://vscode.dev}{vscode.dev}, flowR's extension also works similarly and out of the box, although some configuration options (like switching to an installed R~interpreter) are disabled due to the limitations of the browser environment.

\subsection{Working with the Docker Image}
FlowR's \href{https://github.com/flowr-analysis/flowr/wiki/Interface\#using-the-repl}{read-eval-print loop~(REPL)} is the main way to use the \href{https://hub.docker.com/r/eagleoutice/flowr}{docker image} and explore flowR's features as well as running analyses of real-world projects in an isolated environment. The REPL provides a plethora of commands~--- all starting with a colon~--- to interact with flowR's analysis capabilities. Additionally, the REPL provides auto-completion, a command history, summarized outputs, and REPL-specific configurations.
An example REPL session is shown in \cref{fig:repl-example} below.

\begin{figure}[H]
\newsavebox\ReplBox
\begin{lrbox}{\ReplBox}
\lstfs{8}\def\PromptR{\textcolor{cyan!80!purple!40!blue!62!black}{\textbf{R>}}}
\begin{minted}[escapeinside=!!]{void}
!\textbf{\$}! docker run --rm -it eagleoutice/flowr 
flowR repl using flowR v2.8.6, R grammar v14 !\textcolor{gray}{\sffamily\ldots}!
!\color{gray}use :help to get a list of available commands.!
!\PromptR! !\textbf{:dataflowascii}! x <- 2
 c<2>c----------0<1>0
 |<- |  v<0>v   | 2 |
 c---c--| x |---0---0
        v---v
Edges:
2 :to: 1: reads, argument  2 :to: 0: returns, argument
0 :to: 1: defined-by       0 :to: 2: defined-by
!\PromptR! !\textbf{:quit}!
\end{minted}
\end{lrbox}
\centerline{\kern-2pt\tikz{%
   \node[Rect,gray,outer sep=0pt] {\usebox\ReplBox};
}}\vspace{-.42\baselineskip}
\caption{Example REPL Session}\label{fig:repl-example}
\end{figure}\vspace{-.65\baselineskip}

Please note that while the REPL can access the file system using the \textit{file://} prefix, the corresponding directories have to be \href{https://docs.docker.com/engine/storage/bind-mounts/}{mounted into the docker container} (this is also exemplified in the video).

Besides the REPL, the docker image also supports starting a TCP~server using the \texttt{-\null-server} flag, as well as a websocket server by additionally passing the \texttt{-\null-ws} flag. Consult the \href{https://github.com/flowr-analysis/flowr/wiki/Interface\#communicating-with-the-server}{Interface wiki page} for more information on how to interact with the server instance.

\subsection{Programming with the flowR~API}

All important components of flowR's API are explained in the \href{https://github.com/flowr-analysis/flowr/wiki/Analyzer}{wiki pages} including code-examples and automatically updated hyperlinks into the source code. In the demonstration video, we construct a brief example in which we use the \href{https://github.com/flowr-analysis/flowr/wiki/Query-API}{query~API} to infer the data frame shape of the first argument of all \texttt{print} function calls in a given project:

\xlstsetmintedstyle{plain}
\begingroup\lstfs{8}
\begin{minted}[
   morekeywords={const},
   morekeywords={[2]{FlowrAnalyzerBuilder,setEngine,build,addRequest,dataflow,VertexType,verticesOfType,EmptyArgument,SingleSlicingCriterion,query,log}},
   keywordstyle={[2]\color{darkgray}\itshape}
]{typescript}
const analyzer = await new FlowrAnalyzerBuilder()
      .setEngine('tree-sitter')
      .build() ;
analyzer.addRequest('file://' + folder);

const df = await analyzer.dataflow();
const calls = df.graph
      .verticesOfType(VertexType.FunctionCall)
      .filter(([,v]) => v.name === 'print');
for(const [,v] of calls) {
   const firstArg = v.args[0];
   if(firstArg !== EmptyArgument) {
      const criterion: SingleSlicingCriterion 
               = `$${firstArg.nodeId}`;
      const shape = await analyzer.query([{
         type: 'df-shape',
         criterion
      }])
      console.log(
            shape["df-shape"].domains
               .get(criterion)?.toString()
      )
   }
}
\end{minted}
\endgroup

To ease the usage of the API, we already have sample repositories like \href{https://github.com/flowr-analysis/sample-analyzer-project-query}{flowr-analysis/sample-analyzer-project-query} that can be used as a starting point for building new analyses on top of flowR.

\section{Tool Information}\label{sec:tool-information}
As stated in the paper, flowR is available as an extension for VS~Code and Positron, as a Docker image, and as an npm and R~package.
We distribute the source code alongside its extensive documentation under a \textit{GPL-3.0}~license on GitHub: \url{https://github.com/flowr-analysis/flowr}.
With a test-suite of more then \num{7700} tests, continuous integration, and various benchmarks and evaluation on large real-world datasets (e.g,~\cite{sihler2025statically,DBLP:conf/msr/SihlerPSTDD24}), we consider \textit{flowR} to be mature.
Regarding users we note around \num{1500}~installations of the newest version of our extensions and around~\num{5000}~pulls of the Docker image. The documentation is available at \url{https://github.com/flowr-analysis/flowr/wiki} with a rendered version of the in-source documentation at \url{https://flowr-analysis.github.io/flowr/doc/index.html}.
\end{document}